\documentclass[journal]{IEEEtran}

\usepackage{amsmath,amssymb,amsfonts}
\usepackage{physics}
\usepackage{graphicx}
\usepackage{booktabs}
\usepackage{algorithm}
\usepackage{algpseudocode}
\usepackage{cite}
\usepackage{hyperref}
\usepackage{xcolor}
\usepackage{quantikz}
\usepackage{orcidlink}

\IEEEoverridecommandlockouts

\title{Real Variance-Based Variational Quantum Eigensolver for Non-Hermitian Matrices}

\author{
\IEEEauthorblockA{Durgesh Pandey\orcidlink{0009-0001-0877-1980}$^1$, Ankit Kumar Das\orcidlink{0009-0001-2252-2657}$^1$, and P. Arumugam\orcidlink{0000-0001-9624-8024}$^{1,2}$}
    
\IEEEauthorblockA{$^1$Department of Physics, Indian Institute of Technology Roorkee, Roorkee, Uttarakhand 247667, India}

\IEEEauthorblockA{$^2$Centre for Photonics and Quantum Communication Technology, Indian Institute of Technology Roorkee, Roorkee 247667, India}

\IEEEauthorblockA{Email: durgesh\_p@ph.iitr.ac.in, ankitk\_das@ph.iitr.ac.in, arumugam@ph.iitr.ac.in}
}

\begin{document}
\maketitle

\begin{abstract}
Non-Hermitian operators naturally arise in the description of open quantum systems, which exhibit features such as resonances and decay processes, where the associated eigenvalues are complex. Standard quantum algorithms, including the Variational Quantum Eigensolver (VQE), are designed for Hermitian operators and are ineffective in recovering correct eigenvalues for non-Hermitian matrices. We present a systematic formulation based on a Real Variance-based Variational Quantum Eigensolver (RVVQE) for non-Hermitian operators. A correct cost function that guarantees convergence to the true eigenstates is identified. Our implementation utilizes Hermitian measurements only, rendering the algorithm easily deliverable. The performance and scalability of the proposed algorithm on a hierarchy of dense non-Hermitian matrices of increasing dimension are demonstrated with numerical results and computational metrics.
\end{abstract}

\section{Introduction}
The eigenvalue problem associated with the diagonalization of a matrix is very fundamental in mathematics and physics~\cite{griffiths2018introduction,sakurai2020modern,shankar1994principles}. The memory and other computational resource requirements for diagonalizing a large matrix are a computational challenge that imposes several limits on the classical computer. This classical limit could be stretched further with quantum computing~\cite{preskill2018nisq}. Most quantum algorithms rely on the variational quantum eigensolver (VQE), designed for Hermitian matrices, yielding the least eigenvalue (e.g. ground state energy) and with some extension to yield the other chosen states (e.g. excited states)~\cite{mcclean2016theory, Peruzzo2014VQE, PhysRevC.104.034301}. These algorithms are well-suited for closed-system dynamics, as the Hermitian operators govern such environments.~\cite{mcclean2016theory}. For the non-Hermitian operators of the open system~\cite{Moiseyev2011NonHermitianQM, Rotter2009NonHermitianReview, xie_variational_QA} the well-established VQE is not applicable directly. There are a few algorithms available for these non-Hermitian operators, but they either require extensive parameter space scanning or rely heavily on accurate initial guesses~\cite{quantum_simulation_nonHermitianHamiltonian, Non-Hermitianground-state-searchingalgorithmenhancedbyavariationaltoolbox, xie_variational_QA}. Many of them proceed by indirectly converting non-Hermitian operators into a Hermitian one, which is not always feasible and more problem‑specific~\cite{Moiseyev2011NonHermitianQM, xie_variational_QA, barch2025computationalcomplexitynonhermitianquantum}. The open quantum system is a broad field where the systems interact with the environment and the properties governed by non‑Hermitian operators~\cite{Moiseyev2011NonHermitianQM, Rotter2009NonHermitianReview, Singh2025Quantumsimulationsofnuclearresonanceswithvariationalmethods}. Quantum algorithms capable of efficiently solving not only the non‑Hermitian system but also generating all excited states are crucial for solving many real-world problems~\cite{xie_variational_QA}.

Quantum mechanics traditionally relies on Hermitian operators to represent physical observables, ensuring real eigenvalues and unitary time evolution. This framework is sufficient for closed systems but insufficient for most of the naturally occurring systems. These are the open systems interacting with other systems or an environment~\cite{Moiseyev2011NonHermitianQM,Rotter2009NonHermitianReview}. 
Generally, these effective non-Hermitian operators yield complex eigenvalues of the form $E = E_r - i E_i$. In the context of the Hamiltonian describing an open quantum system, the eigenvalues correspond to the possible resonance states associated with the eigenvalues
\begin{equation}
E = E_r - i\frac{\Gamma}{2},
\end{equation}
where $E_r$ represents the resonance energy and $\Gamma$ denotes the decay or growth rate~\cite{Moiseyev2011NonHermitianQM,Rotter2009NonHermitianReview,xie_variational_QA}. Accurate computation of these complex eigenvalues is essential in several domains, including nuclear physics, quantum chemistry, and scattering theory~\cite{Moiseyev2011NonHermitianQM,xie_variational_QA}. The non-Hermitian description is crucial in applications like large-scale power transmission networks\cite{10746623}, power grid interactions~\cite{Zhang2021_CommPhys}, climate dynamics\cite{Zhang_Xie_2024}, photosynthetic biological processes~\cite{Kassal2013}, mitochondrial energy dissipation in cellular genetics, targeted nanoparticle drug delivery trajectories~\cite{cleve2016efficient, targetted_drug_delievery_Complex}, nanoscale optical microcavity sensing\cite{Chen2017}, to name a few. 

With recent advances in quantum hardware, the emphasis is on casting the eigenvalue problem as a variational one where the expectation values of the operators are efficiently calculated.  The variational quantum algorithms, though popular, are fundamentally incompatible with non-Hermitian operators,  motivating the development of new cost functions and optimization strategies~\cite{mcclean2016theory, Peruzzo2014VQE, zhang2020variational, xie_variational_QA}.  The details of our approach in this regard is presented in the forthcoming sections.

\section{Variance-Based Variational Quantum Eigensolver}
Let $M$ be a general non-Hermitian operator. Its eigenvalue equation,
\begin{equation}
M|\psi\rangle = \lambda |\psi\rangle,
\end{equation}
admits complex eigenvalues and eigenvectors~\cite{Moiseyev2011NonHermitianQM, Rotter2009NonHermitianReview}.
Such a non-Hermitian matrix cannot be measured directly on quantum hardware due to the limitations of current physical apparatus, yielding only real-valued outcomes. However, to overcome this limitation, a cost function that preserves eigenstate identity while allowing decomposition into Hermitian forms is required. A defining property of an eigenstate $|\psi\rangle$ of any operator $M$ is the vanishing of its variance, given by the condition
\begin{equation}
\Delta(M) = \langle M^\dagger M \rangle - |\langle M \rangle|^2 = 0,
\end{equation}
which is both necessary and sufficient~\cite{10.1063/1.1621615,xie_variational_QA}. Therefore, minimizing the variance provides a mathematically rigorous cost function for identifying eigenstates, independent of whether the spectrum is real or complex~\cite{10.1063/1.1621615,xie_variational_QA}. The variance of a non-Hermitian operator $M$ is written as
\begin{equation}
\Delta(M) = \langle M^\dagger M \rangle - |\langle M \rangle|^2,
\end{equation}
which is a strictly real and non-negative quantity~\cite{10.1063/1.1621615,xie_variational_QA}. Using the decomposition
$M = H + iK$, where
\begin{equation}
\label{components_of_M}
H = \frac{M + M^\dagger}{2}, \quad \text{and } 
K = \frac{M - M^\dagger}{2i},
\end{equation}
the variance can be expanded as
\begin{align}
\label{var_in_component_form}
\Delta(M) =
\langle H^2 \rangle - \langle H \rangle^2
+ \langle K^2 \rangle - \langle K \rangle^2 - 
i\langle [H,K] \rangle.
\end{align}

Considering the fact that eigenstates of an operator are simultaneous eigenstates of its Hermitian and anti-Hermitian components, we get the condition for a true eigenstate $|\psi\rangle$.
The expectation value of the commutator $\langle [H,K] \rangle$ will vanish if a state is an eigenstate~\cite{xie_variational_QA}. Consequently, the imaginary term does not contribute to the solution. Since variational optimization requires a real scalar objective function, we define the real part of the variance as a practical cost function. The value is always positive for this cost function with minimum values at zero only, for the eigenstates.

\begin{figure*}[t]
    \centering
    \includegraphics[width=1\linewidth]{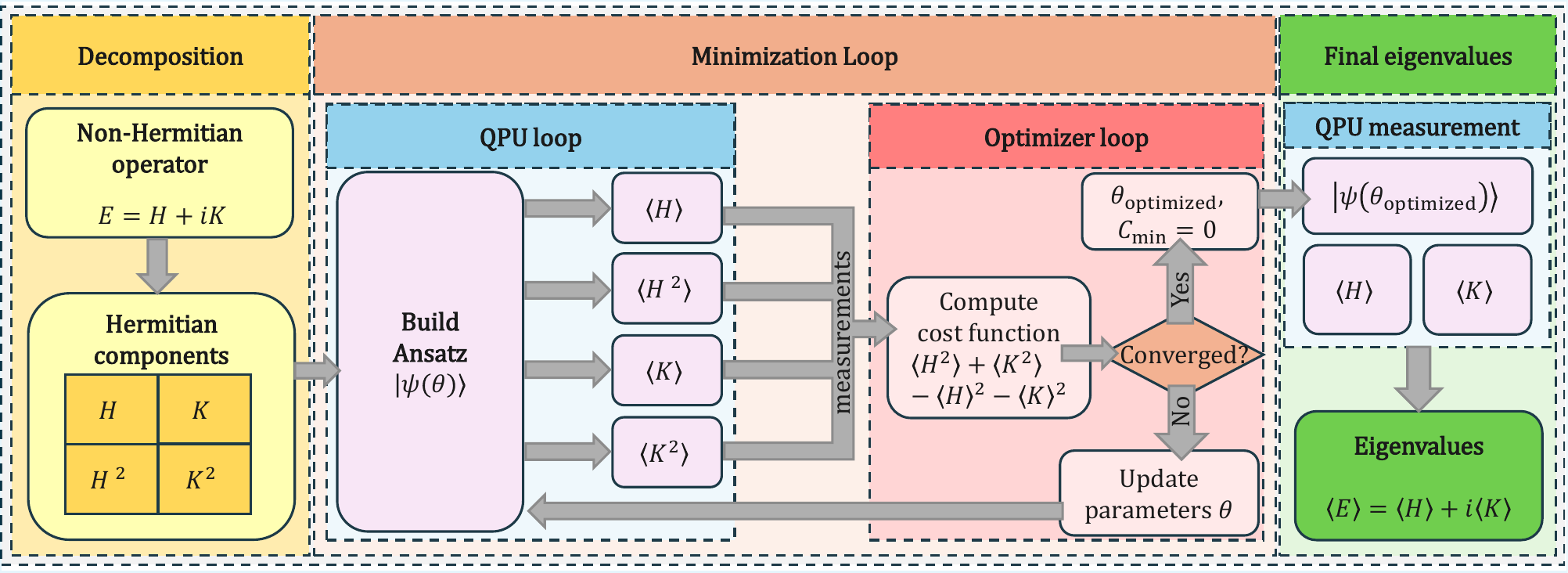}
    \caption{Quantum algorithm to find eigenvalues and eigenstates for non-Hermitian operators}
    \label{fig:vvqe_algorithm_flowchart}
\end{figure*}

The proposed real variance based VQE (RVVQE) cost function is
\begin{equation}
\label{cost_function_RVVQE}
\boxed{
C_{\text{var}}(\theta)= \langle H^2 \rangle_\theta + \langle K^2 \rangle_\theta - (\langle K \rangle_\theta)^2 - (\langle H \rangle_\theta)^2 .
}
\end{equation}
Minimization of $C_{\text{var}}(\theta)$ is sufficient to recover true
eigenstates of $M$, while guaranteeing a real-valued, physically meaningful
cost function compatible with hybrid quantum-classical optimization~\cite{10.1063/1.1621615,xie_variational_QA}. Being Hermitian in nature, the measurement of $H, K, H^2, \text{and } K^2$
components can be done using current quantum hardware. 
Finally,  the eigenvalue  $\langle M \rangle$ corresponds to the eigenstates obtained by minimizing $C_{\text{var}}(\theta)$ is calculated using 
\begin{equation}
\langle M\rangle = \langle H \rangle + i \langle K \rangle.
\end{equation}
Based on the variance minimization principle derived above, we define a hybrid quantum-classical workflow that allows for the extraction of complex spectra using standard quantum hardware.

\section{Algorithm}
\label{algorithm}
The quantum algorithm for obtaining the eigenstates of a non-Hermitian matrix is presented in Algorithm \ref{algorithm_1}. 

\begin{algorithm}[h]
\caption{Quantum Algorithm for RVVQE}
\label{algorithm_1}
\begin{algorithmic}[1] \renewcommand{\algorithmicrequire}{\textbf{Input:}}
 \renewcommand{\algorithmicensure}{\textbf{Output:}}
\Require {Non-Hermitian matrix $M$, initial parameters $\theta$, tolerance $\epsilon$}
\Ensure {Optimized eigenstate $|\psi(\theta^*)\rangle$, eigenvalue $\langle M \rangle$}

\State Decompose $M = H + iK$ where $H, K$ are Hermitian
\State Initialize parametrized quantum state $|\psi(\theta)\rangle$
\Repeat
    \State Measure expectation values 
    \Statex \hspace{1.5em} $\langle H \rangle, \langle K \rangle, \langle H^2 \rangle, \langle K^2 \rangle$
    \State Compute cost function:
    \Statex \hspace{1.5em} $C_{\text{var}}(\theta) = \langle H^2 \rangle_\theta - \langle H \rangle_\theta^2 + \langle K^2 \rangle_\theta - \langle K \rangle_\theta^2$
    \State $\theta \gets \text{Update parameters via classical optimizer}$
\Until{change in $C_{\text{var}}(\theta) < \epsilon$}
\State Identify optimal parameters:$\theta^* \gets \theta$
\State Compute eigenvalue:
\Statex \hspace{1em} $\langle M \rangle = \langle \psi(\theta^*) | H | \psi(\theta^*) \rangle + i \langle \psi(\theta^*) | K | \psi(\theta^*) \rangle$
\State \Return eigenstate $|\psi(\theta^*)\rangle$ and eigenvalue $\langle M \rangle$
\end{algorithmic}

\end{algorithm}
\section{Numerical Results}

We evaluate the performance of the proposed algorithm on a hierarchy of dense non-Hermitian matrices of increasing dimensions. 
The goal here is to demonstrate that minimizing the real-valued variance cost function reliably converges to the true complex eigenvalues, and to assess the scalability of the approach for quantum hardware.

\subsection{Test Matrices}

We consider three dense non-Hermitian matrices: a $2\times2$ matrix ($M_1$), a $4\times4$ matrix ($M_2$), and an $8\times8$ matrix ($M_3$), listed in Appendix~\ref{app1}. 
These matrices have complex spectra and do not fall into the trivial cases where standard Hermitian techniques apply. Exact eigenvalues are obtained using classical diagonalization and used solely as reference values.

\begin{figure*}[t]
    \centering
        \begin{quantikz}
        \lstick{$\ket{0}$} & \qw & \gate{\shortstack{$R_z$\\[-2pt]\scriptsize$\theta_1$}} & \gate{\shortstack{$R_y$\\[-2pt]\scriptsize$\theta_2$}} & \gate{\shortstack{$R_z$\\[-2pt]\scriptsize$\theta_3$}} & \ctrl{1} & \qw & \qw & \qw & \ctrl{2} & \qw & \qw & \qw & \qw & \qw & \qw \\
        \lstick{$\ket{0}$} & \qw & \qw & \qw & \qw & \targ{} & \gate{\shortstack{$R_z$\\[-2pt]\scriptsize$\theta_4$}} & \gate{\shortstack{$R_y$\\[-2pt]\scriptsize$\theta_5$}} & \gate{\shortstack{$R_z$\\[-2pt]\scriptsize$\theta_6$}} & \qw & \ctrl{1} & \qw & \qw & \qw & \qw & \qw\\
        \lstick{$\ket{0}$} & \qw & \qw & \qw & \qw & \qw & \qw & \qw & \qw & \targ{} & \targ{} & \gate{\shortstack{$R_z$\\[-2pt]\scriptsize$\theta_7$}} & \gate{\shortstack{$R_y$\\[-2pt]\scriptsize$\theta_8$}} & \gate{\shortstack{$R_z$\\[-2pt]\scriptsize$\theta_9$}} & \qw & \qw 
        \end{quantikz}
    \caption{Three-qubit unitary rotation entangled ansatz with nearest-neighbor CNOT gates. The unitary rotation is applied to each qubit to allow complete Hilbert space exploration and CNOT gates to entangle those qubits together. This classic hardware-efficient entangling circuit allows maximum expressibility with shallower circuit depth.}
    \label{fig:ansatz_3q_ent}
\end{figure*}
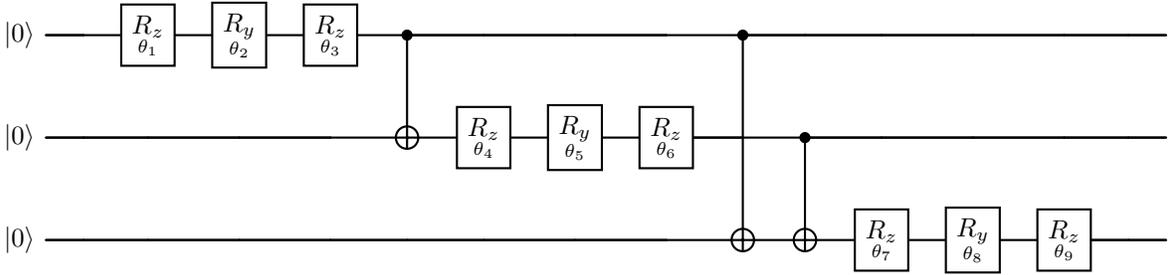

\subsection{Variational Ansatz and Optimization Strategy}

To treat the non-Hermitian nature of $M_n$, we employ a Cartesian decomposition such that $M = H + iK$. Here, both components are Hermitian. This decomposition allows the non-Hermitian problem to be mapped onto a quantum hardware-compatible format by expanding it into a linear combination of Pauli strings. For a matrix of dimension $d$, employ an $n = \lceil\log_2 d \rceil$ qubit variational circuit. For our general purpose of minimizing the cost function via the variational method, we need a trial ansatz with a few parameters. 

The circuit for this ansatz is constructed by comprising the single-qubit Euler rotations $Z-Y-Z$ on each qubit, ensuring full Bloch sphere coverage. This structure is hardware-efficient and ensures full coverage of single-qubit states. To enhance expressibility, we introduce a minimal entangling layer composed of nearest-neighbor CNOT gates. This extension preserves shallow circuit depth while enabling the representation of entangled eigenstates required for accurate variance minimization~\cite{xie_variational_QA, mcclean2016theory}. A simple ansatz for three qubits is as shown in the Fig. \ref{fig:ansatz_3q_ent}.\\

To detect multiple global minima, arising from considering variance as a cost function, we adopt a multi-start optimization strategy. For each matrix, the RVVQE optimization is initialized repeatedly using different sampled initial parameter vectors. We set the number of initial guesses as a multiple of the matrix dimension $d$. Since a $d$-dimensional matrix has at most $d$ eigenvalues, these initial guesses are uniformly distributed for rotation angles from $0$ to $2\pi$, and scaling ensures that the optimizer samples enough distinct regions of the landscape to capture the full complex spectrum. Among all optimization runs, the solution yielding the minimum variance value is obtained, corresponding to all eigenvalues for that matrix.

\subsection{Convergence and Accuracy}

\begin{figure}[h]
    \centering
    \includegraphics[width=1.1\linewidth]{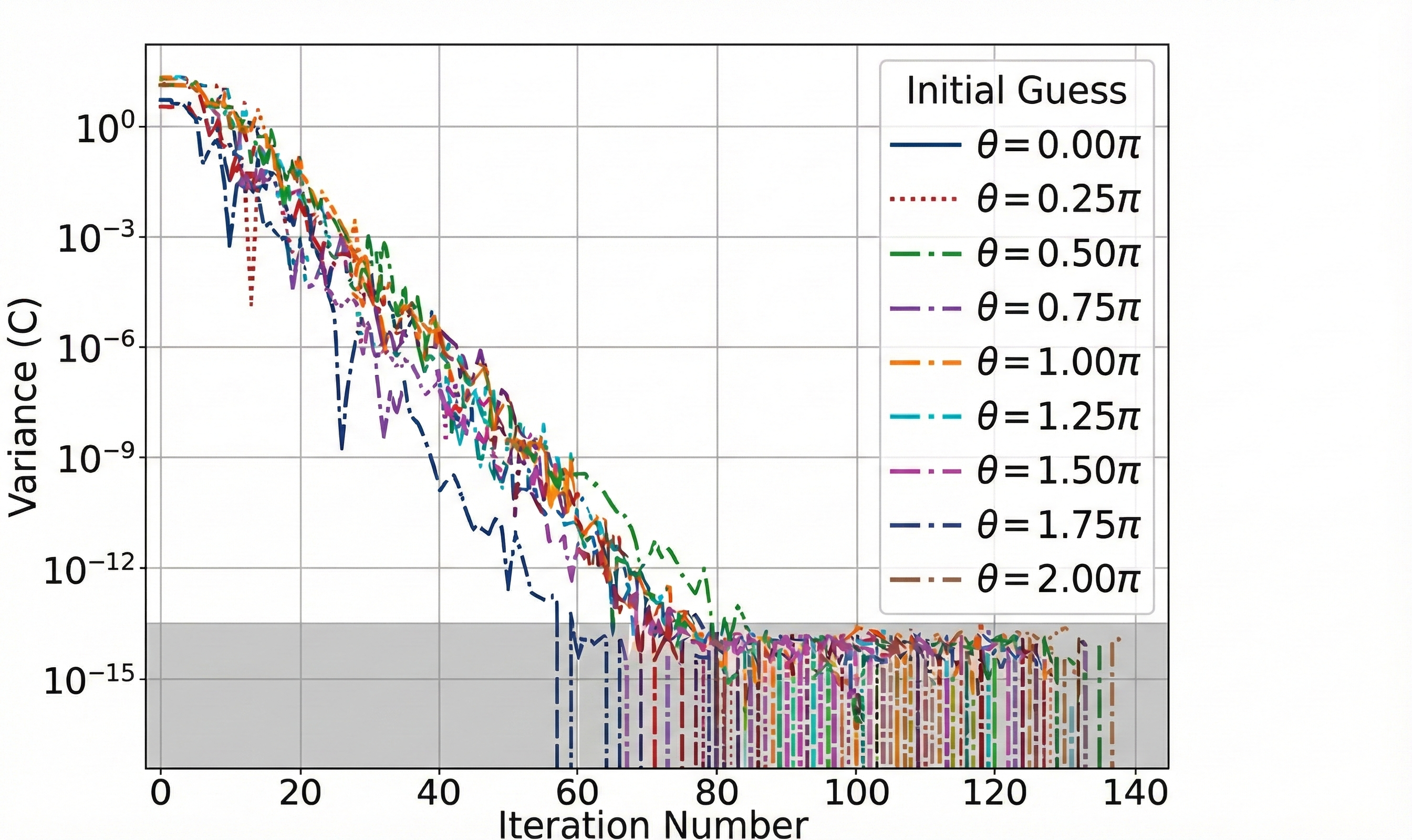}
    \caption{Plot showing convergence of the evaluated cost function from the matrix $M_1$ to zero for different initial guesses of parameters fixed to a single value converging at the computational limit of the system ($10^{-14} - 10^{-16}$, shaded grey).}
    \label{fig:convergence_for_M1}
\end{figure}

Across all tested matrices and initializations, the cost function consistently converges to eigenstates with near-zero variance, indicating successful identification of true eigenvectors. For the $2\times2$ and $4\times4$ cases, the final variance values are typically of the order $\mathcal{O}(10^{-16}- 10^{-9})$, corresponding to machine-precision agreement between the eigenvalues and the exact complex eigenvalues. The Fig.~\ref{fig:convergence_for_M1} shows the value of the real part of the variance converging to zero with iterations for different initial values of parameters fixed to $\theta$.

For the $8\times8$ matrix, the cost function converges consistently. However, it shows a slight degradation in precision, as evidenced by a final variance, with values reaching $\mathcal{O}(10^{-9})$. Despite this increase, the resulting complex eigenvalues match the exact eigenvalues to high numerical accuracy. This behavior reflects the growing expressibility demands placed on the variational ansatz as system size increases, rather than a failure of the RVVQE formulation itself.

An important observation is that different random initializations can converge to distinct eigenvalues of the same non-Hermitian matrix. By systematically scanning multiple initial parameter guesses, RVVQE can recover multiple complex eigenpairs without modifying the cost function. This behavior is consistent with the fact that variance minimization does not privilege a ground state, in contrast to energy-based VQE~\cite{zhang2020variational}.

\subsection{Implications for Quantum Computing}

From a quantum algorithm perspective, these results demonstrate that RVVQE provides a reliable route for accessing complex spectra of non-Hermitian operators using only Hermitian measurements. The use of a layered entangling ansatz is essential for scalability, while the multi-start strategy plays a critical role in navigating the non-convex optimization landscape. Across all tested cases, the real part of the variance converges to eigenvalues that match exact diagonalization within numerical precision, with variances approaching zero up to floating-point error. Corresponding eigenvalues for the finally optimized ansatz at which variance becomes approximately zero are listed in Table~\ref {tab:rvvqe_detailed}. With the increase in the matrix dimension, the minimum achievable variance increases mildly, reflecting the expressibility of the ansatz and optimization complexity rather than algorithmic failure. The convergence of the variance decreases with the increase in parameters for a matrix of increasing dimension, due to the combination of errors from optimizing each of the parameters. This results in absolute error of order $\mathcal{O}(10^{-14}, 10^{-11}, \text{ and } 10^{-9})$ for the $2\cross2, 4\cross4, $ and $8\cross8$ matrices $M_1$, $M_2$, and $M_3$, respectively, despite setting the tolerance of the minimizer to near the computational limit of the order $\mathcal{O}(10^{-15})$. Due to ansatz inexpressibility and the increasing complexity of the optimization landscape (e.g., potential barren plateaus) as system size increases, results possess absolute errors. However, for the less complex system of low dimension, the computational limit of zero variance is achievable.

\vspace{0.8em}

\begin{table}[h]
\centering
\caption{Angle-resolved RVVQE results for non-Hermitian matrices. The reported eigenvalues match exact diagonalization within numerical precision.}
\label{tab:rvvqe_detailed}
\renewcommand{\arraystretch}{1.2}
\begin{tabular}{c c c c}
\toprule
\textbf{Matrix} & \textbf{Angle (rad)} & \textbf{RVVQE Eigenvalue} & \textbf{Variance} \\
\midrule \midrule
$M_1$ & $0.0000$ & $-0.3924 + 0.1050i$ & $\sim 10^{-15}$ \\
$M_1$ & $2.0944$ & $ 5.3924 - 1.1050i$ & $\sim 10^{-15}$ \\
$M_1$ & $4.1888$ & $ 5.3924 - 1.1050i$ & $\sim 10^{-15}$ \\
$M_1$ & $6.2832$ & $-0.3924 + 0.1050i$ & $\sim 10^{-15}$ \\
\midrule
$M_2$ & $0.0000$ & $10.5188 - 0.5892i$ & $\sim 10^{-13}$ \\
$M_2$ & $0.8976$ & $-1.6142 - 0.5359i$ & $\sim 10^{-13}$ \\
$M_2$ & $2.6928$ & $ 3.7513 - 1.0559i$ & $\sim 10^{-13}$ \\
$M_2$ & $3.5904$ & $ 3.3441 + 2.1809i$ & $\sim 10^{-13}$ \\
\midrule
$M_3$ & $0.0000$ & $-0.4866 + 9.6633i$ & $\sim 10^{-9}$ \\
$M_3$ & $0.8976$ & $-0.3361 + 7.4468i$ & $\sim 10^{-9}$ \\
$M_3$ & $2.6928$ & $-0.4866 + 9.6633i$ & $\sim 10^{-9}$ \\
$M_3$ & $3.5904$ & $-0.8231 + 11.8959i$ & $\sim 10^{-9}$ \\
\bottomrule
\end{tabular}
\end{table}

\begin{figure}[h]
    \centering
    \includegraphics[width=1\linewidth]{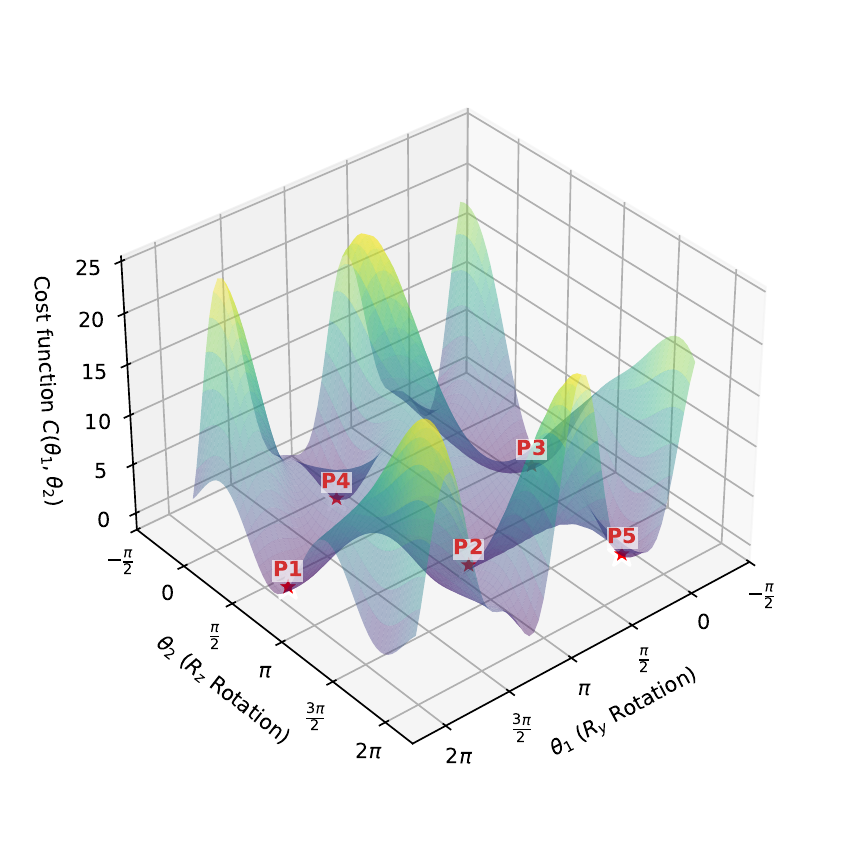}
    \caption{Plot showing real part of the variance as cost function, for the matrix $M_1$ whose ansatz depends on two parameters $\theta_1$ and $\theta_2$. The eigenvalues corresponding to those eigenstates are obtained at the points $P_1$ to $P_5$ at which the cost function is zero.}
    \label{fig:var_nonhermitian}
\end{figure}

Fig. \ref{fig:var_nonhermitian} illustrates the optimization landscape for the variational quantum eigensolver. The distinct valleys in the plot correspond to regions where the quantum state approaches an eigenstate of the non-Hermitian matrix $M_1$. We utilized our hybrid classical-quantum optimization loop to identify the exact minima. Table \ref{tab:optimization_results} details the specific parameters and resulting complex eigenvalues for the five solution points ($P_1$ to $P_5$) identified in Fig.~\ref{fig:var_nonhermitian}. This confirms that the cost function successfully identifies all eigenstates of the non-Hermitian system despite the complex spectrum.

\begin{table}[h]
    \centering
    \caption{Optimized parameters and corresponding eigenvalues for the non-Hermitian Matrix.}
    \label{tab:optimization_results}
    \begin{tabular}{cccc}
        \toprule
        \textbf{Point} & $\boldsymbol{\theta_1}$ \textbf{(rad)} & $\boldsymbol{\theta_2}$ \textbf{(rad)} & \textbf{Eigenvalue} $\boldsymbol{\lambda}$ \\
        \midrule \midrule
        $P_{1}$ & $6.2467$ & $2.2802$ & $5.3924 - 1.1050i$ \\
        $P_{2}$ & $3.1051$ & $4.0030$ & $5.3924 - 1.1050i$ \\
        $P_{3}$ & $-0.0365$ & $2.2802$ & $5.3924 - 1.1050i$ \\
        $P_{4}$ & $3.7084$ & $0.6173$ & $-0.3924 + 0.1050i$ \\
        $P_{5}$ & $0.5668$ & $5.6658$ & $-0.3924 + 0.1050i$ \\
        \bottomrule
    \end{tabular}
\end{table}

\section{Wider Scope of RVVQE}

The method to obtain eigenvalues using the mentioned RVVQE acts as a generalized implementation framework for any kind of matrix. We applied this to a broad class of $2\times2$ matrices, encompassing real symmetric, real antisymmetric, complex non-Hermitian, and purely imaginary cases. This setting provides a controlled testbed to validate the robustness of the proposed cost function and optimization strategy before scaling to higher-dimensional systems. We consider six matrices labelled \(A\) to \(F\), where \(A\) is real symmetric, \(B\) is real skew-symmetric, \(C\) is real non-symmetric, \(D\) is complex Hermitian, \(E\) is complex skew-Hermitian, and \(F\) is a general complex non-Hermitian matrix, mentioned in Appendix~\ref{app1}.


The algorithm employs a single-qubit parametrized ansatz composed of Euler rotations,
\begin{equation}
U(\boldsymbol{\theta}) = R_z(\theta_1) R_y(\theta_2) R_z(\theta_3),
\end{equation}
acting on an initial computational basis state. The cost function is defined as mentioned in Sec.~\ref{algorithm}, which vanishes when the trial state converges to an eigenstate.  For each matrix, the optimization is repeated for multiple initial parameter guesses and for a discrete sweep of an external angle parameter $\theta \in [0,2\pi]$.


Across all tested matrices, the RVVQE consistently converges to exact eigenvalues within numerical precision, with variance values typically in the range $10^{-15}-10^{-14}$. Real symmetric matrices exhibit purely real eigenvalues, while antisymmetric and complex matrices correctly yield imaginary or fully complex spectra. The observed switching between distinct eigenvalues with initial guess of $\theta$ reflects the multi-directional converging of the variance cost function, confirming that the algorithm is capable of finding different global minima to compute all the possible eigenstates.

\begin{table}[h]
\centering
\caption{Eigenvalues obtained for different initial guess for parameter $\theta$ for assumed $2\times2$ matrices.}
\label{tab:rvvqe_2x2_table}
\renewcommand{\arraystretch}{1.2}
\begin{tabular}{c c c}
\hline
\textbf{Matrix} & $\boldsymbol{\theta}$\textbf{(rad)} & \textbf{RVVQE Eigenvalues} \\
\hline \hline
$A$ & $0$ & $5.8541 + 0.0000i$ \\
\quad & $\pi$ & $-0.8541 + 0.0000i$ \\
\hline
$B$ & $0$ & $0.0000-2.0000 i$ \\
\quad & $\pi/2$ & $0.0000+2.0000 i$ \\
\hline
$C$ & $0$ & $5.3723 + 0.0000i$ \\
\quad & $\pi$ & $-0.3723 + 0.0000i$ \\
\hline
$D$ & $0$ & $4.4495 + 0.0000i$ \\
\quad & $\pi/2$ & $-0.4495 + 0.0000i$ \\
\hline
$E$ & $0$ & $0.0000+3.0000 i$ \\
\quad & $\pi$ & $0.0000 + 0.0000 i$ \\
\hline
$F$ & $0$ & $5.3924 - 1.1050 i$ \\
\quad & $\pi/3$ & $-0.3924 + 0.1050 i$ \\
\hline
\end{tabular}
\label{rvvqe for all matrices}
\end{table}
The results in the Table~\ref{rvvqe for all matrices} demonstrate that the proposed RVVQE formulation is independent of matrix structure and accurately captures real, imaginary, and complex eigenvalues using a single-qubit parametrization.

\subsection{Computational Comparison: Hermitian Case ($M=M^\dagger$)}

For Hermitian systems, there is no anti-Hermitian part $K$ as in Equations~\ref{components_of_M} and \ref{var_in_component_form}, resulting in theoretical equivalence of RVVQE to the VVQE .The choice between Standard VQE and RVVQE presents a significant trade-off between computational efficiency and spectral capability.

Standard VQE scales linearly with the system size because it only involves measuring the Hamiltonian $H$. In contrast, RVVQE minimizes variance, which requires evaluating $\langle H^2 \rangle$. Squaring the Hamiltonian causes the number of Pauli terms to scale quadratically ($\mathcal{O}(N_\text{terms}^2)$), leading to a significantly higher measurement overhead (shot cost) per optimization step.

\begin{figure}
    \centering
    \includegraphics[width=1\linewidth]{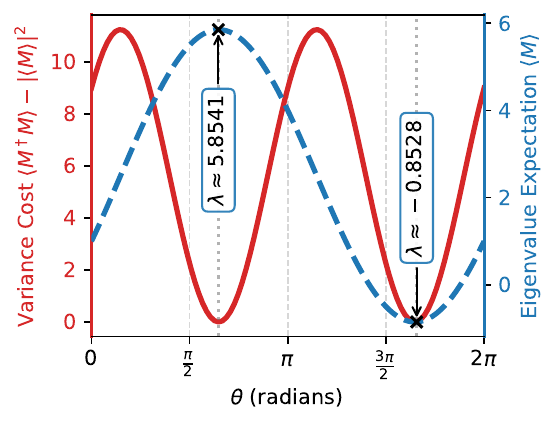}
    \caption{Plot showing correlation between the eigenvalue for matrix $A$ (dashed blue line) and real part of the variance as cost function (solid red line), whose ansatz depends only on a single parameter $\theta$. The maxima and minima of the eigenvalue expectations lie exactly on the global minima, which is zero for those particular values of the parameter $\theta$}
    \label{fig:eigenvalue_vs_costfunction}
\end{figure}

Depending upon the difference in the optimization Landscape, the runtime is accordingly affected. Standard VQE typically presents a valley landscape that naturally guides the optimizer toward the ground state. RVVQE, however, creates a landscape where every eigenstate is a global minimum (zero variance), as shown in Fig. \ref{fig:eigenvalue_vs_costfunction} for the matrix $A$. While this allows access to excited states, it makes the optimization surface more complex, often requiring more iterations to escape barren plateaus and converge.

\begin{figure}
    \centering
    \includegraphics[width=1\linewidth]{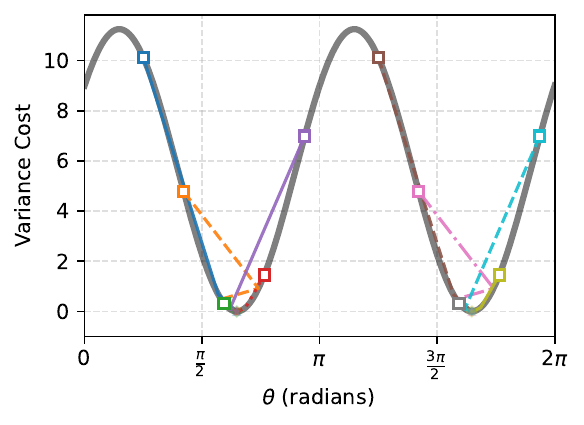}
    \caption{The solid curve represents the cost landscape, while the squares denote ten distinct initial starting points for the optimization process. The plot illustrates the convergence behavior of the optimizer from different regions of the ansatz parameterization toward nearest global minima.}
    \label{fig:convergence}
\end{figure}

\begin{table}[h!]
    \centering
    \caption{Comparison of computational metrics.}
    \label{tab:comp_cost_clean}
    \renewcommand{\arraystretch}{1.2}
    \setlength{\tabcolsep}{10pt}
    \begin{tabular}{p{1.5 cm}p{2.0cm}p{3.0cm}}
    \toprule
    \textbf{Metric} & \textbf{Standard VQE} & \textbf{Real Variance - VQE} \quad \textbf{(Algorithm proposed)} \\
    \midrule
    \midrule 
    Primary Goal & Ground State only & Entire Spectrum (Ground with Excited States) \\
    \midrule
    Objective Func. & $\min \langle M \rangle$ & $\min (\langle M^\dagger M \rangle - |\langle M \rangle|^2)$ \\
    \midrule
    Scaling & Linear $\mathcal{O}(N_\text{terms})$ & atmost Quadratic $\mathcal{O}(N_\text{terms}^2)$ \\
    \midrule
    Applicability & Only Hermitian matrices & Also for non-Hermitian matrices \\
    \midrule
    Measurement & Low cost (measures $H$) & High cost (measures real part of $H^2$ which consists of 3 different measuremnts) \\
    \midrule
    Convergence & Fast (Simple landscape) & Slower (Complex landscape) \\
    \midrule
    Convergence value & Unknown & Zero \\
    \bottomrule
    \end{tabular}
\end{table}

Table \ref{tab:comp_cost_clean} summarizes the distinct operational differences between the algorithms VQE and RVVQE. This clearly states that the standard VQE is computationally superior for finding ground states as it does not require calculating $\langle H^2 \rangle$. RVVQE is strictly necessary only when the system is non-Hermitian or when specific excited states are required. For Hermitian ground states, standard VQE is superior due to linear scaling $\mathcal{O}(N_\text{terms})$, whereas RVVQE's quadratic scaling $\mathcal{O}(N_\text{terms}^2)$ is only justified when complex eigenvalues or non-Hermitian dynamics are involved.

\section{Conclusion}

We presented a simple and direct variational framework for solving non-Hermitian eigenvalue problems using variance minimization. Our results establish RVVQE as a bridge for simulating open quantum systems on NISQ hardware, providing a method for extracting complex eigenvalues without the need for non-unitary gate operations. The approach overcomes fundamental failures of expectation-value-based methods and guarantees convergence to true eigenstates on present quantum hardware. Our results establish the real part of the variance minimization as the correct principle for extending variational quantum algorithms to open quantum systems and complex spectra. This is a practical solution until another method for mapping open quantum systems directly onto quantum hardware, such as linking decoherence to the imaginary part of the eigenvalue, is found.

\section{acknowledgements}
This work is supported by the SERB-DST, Govt.~of India, via project \sloppy{CRG/2022/009359}.
We acknowledge the National Supercomputing Mission (NSM) for providing computing resources of `PARAM Ganga' at IIT Roorkee, which is implemented by C-DAC and supported by MeitY and DST, Govt.~of India.


\appendix
\section*{Chosen matrices}
\label{app1}
We consider the following matrices $A$ through $F$, categorizing them by their properties, alongside larger non-Hermitian matrices $M_1, M_2,$ and $M_3$.

\begin{table}[h!]
\centering
\renewcommand{\arraystretch}{2}
\begin{tabular}{|p{1.5cm}|p{3.9cm}|}
\hline
\textbf{Matrix Type} & \textbf{Matrix Definition} \\ \hline
Real Symmetric & $A = \begin{pmatrix} 1 & 3 \\ 3 & 4 \end{pmatrix}$ \\ \hline
Real Skew-Symmetric & $B = \begin{pmatrix} 0 & 2 \\ -2 & 0 \end{pmatrix}$ \\ \hline
Real Non-Symmetric & $C = \begin{pmatrix} 1 & 2 \\ 3 & 4 \end{pmatrix}$ \\ \hline
Complex Hermitian & $D = \begin{pmatrix} 1 & 2+i \\ 2-i & 3 \end{pmatrix}$ \\ \hline
Complex Skew-Hermitian & $E = \begin{pmatrix} i & 1-i \\ -1-i & 2i \end{pmatrix}$ \\ \hline
Complex Non-Hermitian & $F = M_1 = \begin{pmatrix} 1+i & 2-i \\ 3+2i & 4-2i \end{pmatrix}$ \\ \hline
\end{tabular}
\end{table}

Larger non-Hermitian matrices $M_2$ and $M_3$ :

\[
M_2 = \begin{pmatrix} 1+i & 2-i & 1+2i & 3-i \\ 3+2i & 4-2i & 2+i & 1-i \\ 1-i & 3+i & 5+2i & 2-2i \\ 2+2i & 1-3i & 4+i & 6-i \end{pmatrix}
\]

\[
M_3 = \begin{pmatrix} 1+i & 1-2i & 1-3i & \cdots & 1-8i \\ 2-i & 2+2i & 2-3i & \cdots & 2-8i \\ 3-i & 3-2i & 3+3i & \cdots & 3-8i \\ \vdots & \vdots & \vdots & \ddots & \vdots \\ 8-i & 8-2i & 8-3i & \cdots & 8+8i \end{pmatrix}
\]

\end{document}